\newcommand{\be}{\begin{equation}}
\newcommand{\ee}{\end{equation}}
\newcommand{\ba}{\begin{eqnarray}}
\newcommand{\ea}{\end{eqnarray}}

\newcommand{\WMAP}    {{\sl WMAP}}

\documentclass[numberedappendix,letter]{emulateapj}
\usepackage{graphicx}
\usepackage{bm}
\usepackage{natbib}
\usepackage{mathrsfs}
\usepackage{color}
\usepackage[mediumspace,Gray,squaren]{SIunits}

\shorttitle{Clustering template}
\shortauthors{G.~Addison et al.}

\begin{document}

\title{Power-law template for IR point source clustering}

\author{
 Graeme~E.~Addison\altaffilmark{1},
 Joanna~Dunkley\altaffilmark{1,2,3},
 Amir~Hajian\altaffilmark{4,3,2},
 Marco~Viero\altaffilmark{5,6},
 J.~Richard~Bond\altaffilmark{4},
 Sudeep~Das\altaffilmark{7,2,3},
 Mark~J.~Devlin\altaffilmark{8},
 Mark~Halpern\altaffilmark{9},
 Adam~D.~Hincks\altaffilmark{2},
 Ren\'ee~Hlozek\altaffilmark{3},
 Tobias~A.~Marriage\altaffilmark{10,3},
 Kavilan~Moodley\altaffilmark{11},
 Lyman~A.~Page\altaffilmark{2},
 Erik~D.~Reese\altaffilmark{8},
 Douglas~Scott\altaffilmark{9},
 David~N.~Spergel\altaffilmark{3},
 Suzanne~T.~Staggs\altaffilmark{2},
 Edward~Wollack\altaffilmark{12}
 }
\altaffiltext{1}{Sub-department of Astrophysics, University of Oxford, Denys Wilkinson Building, Keble Road, Oxford OX1 3RH, UK}
\altaffiltext{2}{Joseph Henry Laboratories of Physics, Jadwin Hall,
Princeton University, Princeton, NJ, USA 08544}
\altaffiltext{3}{Department of Astrophysical Sciences, Peyton Hall, 
Princeton University, Princeton, NJ USA 08544}
\altaffiltext{4}{Canadian Institute for Theoretical Astrophysics, University of
Toronto, Toronto, ON, Canada M5S 3H8}
\altaffiltext{5}{California Institute of Technology, 1200 E. California Blvd., Pasadena, CA 91125, USA}
\altaffiltext{6}{Department of Astronomy \& Astrophysics, University of Toronto, 50 St. George Street, Toronto, ON M5S 3H4, Canada}
\altaffiltext{7}{Berkeley Center for Cosmological Physics, LBL and Department of Physics, University of California, Berkeley, CA 94720, USA}
\altaffiltext{8}{Department of Physics and Astronomy, University of Pennsylvania, 209 South 33rd Street, Philadelphia, PA, USA 19104}
\altaffiltext{9}{Department of Physics and Astronomy, University of British Columbia, Vancouver, BC, Canada V6T 1Z4}
\altaffiltext{10}{Dept. of Physics and Astronomy, Johns Hopkins University, 3400 N. Charles St., Baltimore, MD 21218-2686}
\altaffiltext{11}{Astrophysics and Cosmology Research Unit, School of Mathematical Sciences, University of KwaZulu-Natal, Durban, 4041, South Africa}
\altaffiltext{12}{Code 665, NASA/Goddard Space Flight Center,
Greenbelt, MD, USA 20771}

\begin{abstract}
\setlength{\parindent}{0pt} 
We perform a combined fit to angular power spectra of unresolved infrared (IR) point sources from the \emph{Planck} satellite (at 217, 353, 545 and 857~GHz, over angular scales $100\lesssim\ell\lesssim2200$), the Balloon-borne Large-Aperture Submillimeter Telescope (BLAST; 250, 350 and 500~$\mu$m; $1000\lesssim\ell\lesssim9000$), and from correlating BLAST and Atacama Cosmology Telescope (ACT; 148 and 218~GHz) maps. We find that the clustered power over the range of angular scales and frequencies considered is well fit by a simple power law of the form $C_{\ell}^{\textrm{clust}}\propto\ell^{-n}$ with $n=1.25\pm0.06$. While the IR sources are understood to lie at a range of redshifts, with a variety of dust properties, we find that the frequency dependence of the clustering power can be described by the square of a modified blackbody, $\nu^{\beta}B(\nu,T_{\textrm{eff}})$, with a single emissivity index $\beta=2.20\pm0.07$ and effective temperature $T_{\textrm{eff}}=9.7$~$\textrm{K}$. Our predictions for the clustering amplitude are consistent with existing ACT and South Pole Telescope results at around 150 and 220~GHz, as is our prediction for the effective dust spectral index, which we find to be $\alpha_{150-220}=3.68\pm0.07$ between 150 and 220~GHz. Our constraints on the clustering shape and frequency dependence can be used to model the IR clustering as a contaminant in Cosmic Microwave Background anisotropy measurements. The  combined \emph{Planck} and BLAST data also rule out a linear bias clustering model.
\end{abstract}


\section{Introduction}
The angular power spectrum of Cosmic Microwave Background (CMB) temperature fluctuations currently provides vital constraints on cosmological models \citep[e.g.,][]{komatsu/etal:2011,larson/etal:2011}. Experiments including the Atacama Cosmology Telescope (ACT), South Pole Telescope (SPT), and \emph{Planck} satellite are now probing the CMB temperature power spectrum on arcminute scales \citep{das/etal:2011,keisler/etal:2011,ade/etal:2011b}. An improved measurement of the Silk damping tail \citep{silk:1968} improves constraints on, for instance, the scale dependence of primordial fluctuations, important for testing inflationary models, the number of relativistic species, and early-universe exotica \citep[e.g.,][]{komatsu/etal:2011}. On arcminute scales the contribution to the angular power spectrum from the primary CMB fluctuations becomes subdominant to extragalactic foregrounds including infrared and radio point sources \citep[e.g.,][]{white/majumdar:2003,righi/etal:2008}, and the thermal and kinetic Sunyaev Zel'dovich effects \citep[SZ;][]{sunyaev/zeldovich:1970}. Extracting the CMB signal requires understanding how the contribution from these components varies with frequency and angular scale.

The infrared (IR) point source foreground is understood to originate from high-redshift ($z\sim1\!-\!4$) star-forming galaxies whose rest-frame emission peaks in the far-infrared due to thermal emission from dust grains illuminated by starlight \citep[e.g.,][]{bond/etal:1986,bond/etal:1991c,hughes/etal:1998,blain/etal:1999,draine:2003}. While our work concerns observations made in the mm and sub-mm we refer to these dusty sources throughout as `IR' sources. Thermal dust emission from star-forming galaxies is also an important component of the cosmic infrared background \citep[CIB -- e.g.,][]{puget/etal:1996}. Galaxies trace the large-scale structure and so are clustered, with a scale-dependent contribution to the power spectrum in addition to Poisson shot-noise \citep{peebles:1980}. Clustering of IR sources was therefore expected (e.g., \citealt{bond:1996}, and references therein; \citealt{negrello/etal:2007}) and has been detected in \emph{Spitzer Space Telescope} data at 160~$\mu$m \citep{lagache/etal:2007}, by the Balloon-borne Large-Aperture Submillimeter Telescope (BLAST) and \emph{Herschel Space Observatory} at 250, 350 and 500~$\mu$m \citep{viero/etal:2009,cooray/etal:2010,amblard/etal:2011}, in the microwave sky at around 150 and 220~GHz by SPT and ACT \citep{hall/etal:2010,dunkley/etal:2011,shirokoff/etal:2011}, and in early data from \emph{Planck} \citep[hereafter P11]{ade/etal:2011}. Correlations between clustering at different frequencies have also been detected: \citet[hereafter H12]{hajian/etal:2012} measure significant levels of correlation between BLAST maps and ACT maps at 148 and 218~GHz, detecting a clustered component at 4$\sigma$. P11 also find a significant correlation between the \emph{Planck} 217~GHz maps and those at higher frequencies (353, 545 and 857~GHz). Studying the clustered power of dusty galaxies in the sub-mm is simpler than at the mm CMB bands because they are the dominant extragalactic signal. Knowing that significant overlap exists between the sub-mm galaxy population and those sources responsible for the IR foreground in the CMB bands suggests that information about the former can help our understanding of the latter.

In this work we combine large- and small-scale power spectra from \emph{Planck}, BLAST, and correlations between BLAST and ACT to estimate the amplitude and scale dependence of the angular power spectrum of clustered IR sources. We present a simple power-law template to model the clustered source contribution that may be marginalized over when estimating cosmological parameters from foreground-contaminated CMB maps as in, for instance, \cite{hall/etal:2010}, \cite{dunkley/etal:2011}, and \cite{keisler/etal:2011}.

Previous ACT and SPT results \citep{dunkley/etal:2011,shirokoff/etal:2011} have found that the parameters extracted from their CMB spectra are not particularly dependent on the model adopted for the IR clustered power \citep[see also][]{sehgal/etal:2010,fowler/etal:2010}. The ACT and SPT data sets are not yet complete; the final data will include more sky coverage as well as measurements from additional frequency channels. \cite{millea/etal:2012} find that modeling the IR point source clustering incorrectly for the final combined \emph{Planck} and ACT / SPT data sets could introduce a significant bias in cosmological parameters (they estimate 1$\sigma$ based on the discrepancy between two different IR clustering models). It is therefore important to understand the scale and frequency dependence of the clustered power in preparation for this future analysis. Improving constraints on IR clustering will also help constrain the SZ power spectrum.

In this work we are primarily concerned with the IR sources as a contaminant in CMB maps. While the physical properties of this high-redshift star-forming population are important for understanding the star formation history and galaxy evolution, in this analysis we do not attempt to extract information about, for example, the redshift distribution of the sources or the dark matter halos they occupy.

In Section 2 we describe the data we use for our fitting, in Section 3 we explain our assumptions and methods, results are presented in Section 4 and a conclusion follows in Section 5.

\section{Data}

Throughout this work we use `auto-spectrum' to refer to a power spectrum calculated by correlating two maps at the same frequency, and `cross-spectrum' to refer to a spectrum calculated by cross-correlating maps at different frequencies. `BLAST $\times$ ACT' is to be understood to refer to the BLAST / ACT cross-spectra and so on.

We use the BLAST 250, 350 and 500~$\mu$m auto-spectra, 250 $\times$ 350, 250 $\times$ 500 and 350 $\times$ 500~$\mu$m cross-spectra, and BLAST 250, 350 and 500~$\mu$m $\times$ ACT 148 and 218~GHz cross-spectra from H12, and \emph{Planck} 857, 545, 353 and 217~GHz auto-spectra from P11 to construct the template. The ACT data are from the 2008 observing season and the BLAST $\times$ ACT spectra were calculated from the $\sim$8.6 deg$^2$ common to both sets of maps, as described in H12. We take the quadrature sum of the statistical and beam systematic uncertainties in Table 4 of P11 as the error on each \emph{Planck} data point, neglecting any possible correlation in the beam uncertainty across different angular scales. We find that allowing such a correlation has minimal effect on our results (Section 4.1.3). The BLAST and ACT beam uncertainties are subdominant to the statistical (noise) uncertainty across the range of angular scales covered by the BLAST and BLAST $\times$ ACT data and so we likewise neglect any correlation in the errors on these spectra. The temperature to flux conversion factors given in Table 4 of P11 assume a source SED that varies as $I(\nu)\propto\nu^{-1}$; a correction is then applied to convert to the real flux units \citep[see Sections 5.5 of P11 and 7.4.2 of][]{planckHFI:2011}.

We subtract the Galactic dust emission (cirrus) component in the BLAST and BLAST $\times$ ACT spectra as described in  Section 4.3 of H12, assuming the cirrus contribution varies with angular scale as $\ell^{-2.7}$. These spectra are not very sensitive to the details of the cirrus treatment, since they are from a relatively cirrus-free patch of sky \citep[see also][]{das/etal:2011}. We assume that any contribution to the power spectra other than that from IR galaxies (for instance, radio galaxy-IR or SZ-IR correlations) is negligible. Removal of power in the \emph{Planck} spectra that is not from extragalactic IR point sources is described in Sections 2 and 3 of P11.

The \emph{Planck} and BLAST auto-spectra are presented in Figure 1. We also show the 218~GHz ACT power spectrum from \cite{das/etal:2011} and the 220~GHz SPT spectrum from \cite{shirokoff/etal:2011}. A \WMAP-7 best-fit $\Lambda$CDM CMB power spectrum \citep{komatsu/etal:2011} has been subtracted from these points;  we have not subtracted any power from radio point sources or the kinetic Sunyaev-Zel'dovich effect, as these contributions are likely to be subdominant \citep{dunkley/etal:2011,shirokoff/etal:2011}. All the spectra show similar angular scale dependence despite spanning a broad range of frequencies and almost two decades in angular scale. Note that a color correction to account for the different bandpass filters is required to make the Planck and BLAST spectra directly comparable - see Section 4 for details.

\begin{figure*}[t]
	\centering
	\includegraphics{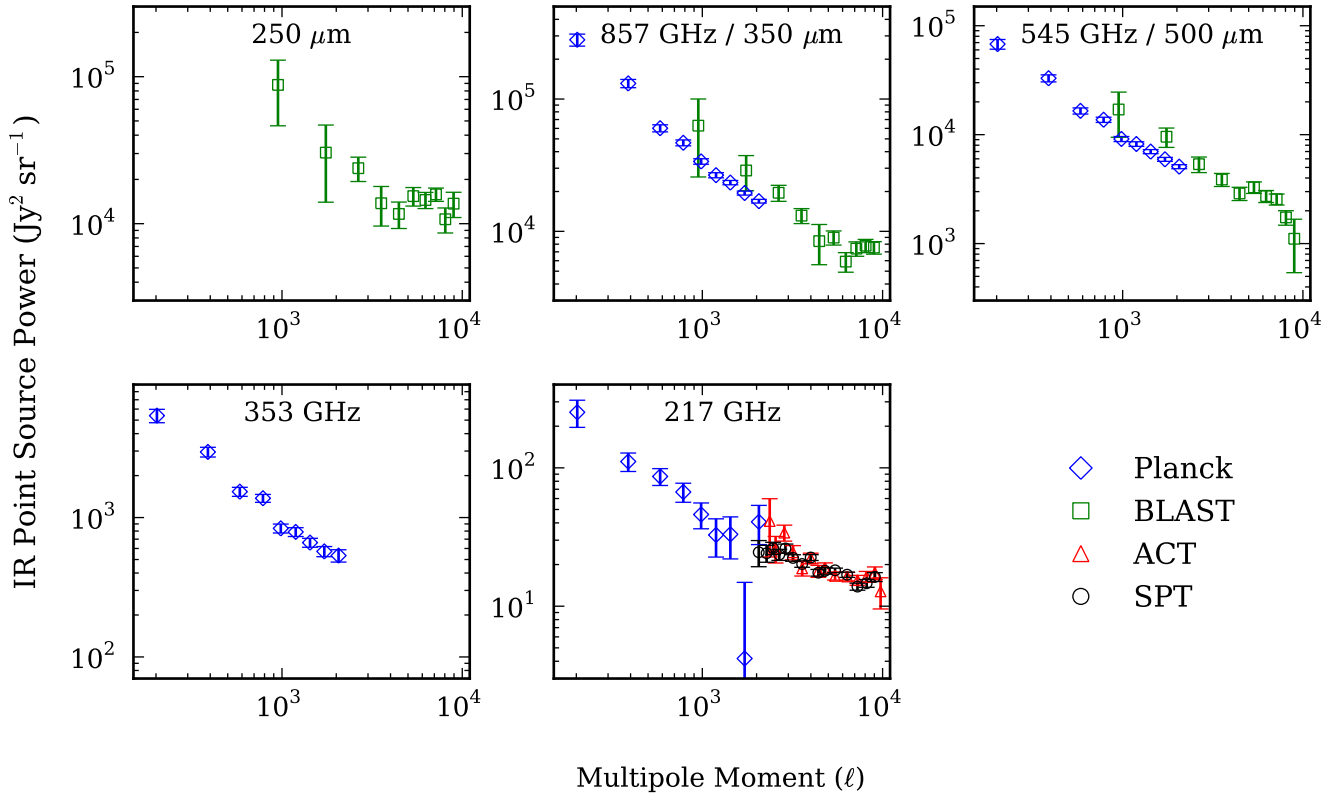}
	\caption{\emph{Planck} (857, 545, 353 and 217~GHz) and BLAST (250, 350 and 500~$\mu$m) IR point source power spectra. The spectra include both shot-noise and clustered components. While these data span a broad range of frequency and angular scale there is a notable similarity in the angular scale dependence. The \emph{Planck} error bars include both the statistical uncertainties and estimates of the systematic beam uncertainty given in Table 4 of \cite{ade/etal:2011}. Data points  at $\ell\gtrsim2000$ from ACT at 218~GHz \citep{das/etal:2011} and SPT at 220~GHz \citep{shirokoff/etal:2011} have been included for comparison with the \emph{Planck} 217~GHz spectrum. We have subtracted a \WMAP-7 best-fit $\Lambda$CDM CMB component from these spectra. No corrections for the different bandpass filter profiles have been applied; when the different filters and photometric calibration uncertainties are accounted for the \emph{Planck} and BLAST data are in good agreement (Section 4).}
\end{figure*}

The \emph{Planck}, BLAST and ACT maps are calibrated using comparisons to various measurements including the orbital CMB dipole, FIRAS data and planetary temperature. Uncertainty in these calibrations must be accounted for in order to perform joint fits to the power spectra extracted from the different maps. The absolute photometric calibration uncertainties are 7\% for the \emph{Planck} 857 and 545~GHz maps, 2\% for the \emph{Planck} 353 and 217~GHz maps \citep[see P11 and][]{planckHFI:2011}, 9.5, 8.7 and 9.2\% for the BLAST 250, 350 and 500~$\mu$m maps \citep{truch/etal:2009}, 7\% for the ACT 218~GHz map and 2\% for the ACT 148~GHz map \citep{hajian/etal:2010}. Furthermore, the uncertainties in the BLAST calibrations are highly correlated; this is because the dominant source of uncertainty is the spectral energy distribution (SED) of the star used for the calibration of all three bands \citep{truch/etal:2009}.

\section{Power-law clustering template}

In the halo model formalism (e.g., \citealt{peacock/smith:2000}; \citealt{scoccimarro/etal:2001}; \citealt{cooray/sheth:2002}; see also earlier work, e.g., \citealt{bond:1996} and references therein), the two-point function of  galaxy clustering is dominated on large scales by pairs of sources in different dark matter halos (the `two-halo' term -- roughly corresponding to the linear clustering regime), while on small scales galaxy pairs occupying the same halo (`one-halo' term -- non-linear regime) become dominant. Although these two components have different scale dependence, an angular correlation function varying as $w(\theta)\propto\theta^{-\delta}$ with $\delta\sim0.8$ \citep[corresponding to clustering power $C_{\ell}^{\textrm{clust}}\propto\ell^{-1.2}$]{peebles:1980} has been found to adequately describe the clustering of, for instance, Lyman Break galaxies at $z\sim3$ \citep{giavalisco/etal:1998} and local SDSS galaxies \citep{zehavi/etal:2002}. Significant deviations from power-law behavior have however been observed in analyses of more recent and comprehensive SDSS data \citep[e.g.,][]{zehavi/etal:2004,zehavi/etal:2005,blake/etal:2008}, as well as in high-redshift galaxies \citep{ouchi/etal:2005,lee/etal:2006,coil/etal:2006,wake/etal:2011}. For a recent discussion of the physical origins of power-law galaxy correlation functions see \cite{watson/etal:2011}.

P11 and \cite{amblard/etal:2011} find that a power law provides an adequate fit to the existing \emph{Planck} and \emph{Herschel}/SPIRE unresolved IR point source spectra (see Tables 6 and S1 in those papers, respectively). This would suggest that IR spectra are not yet of sufficient quality to reveal deviations from power-law clustering behavior and we therefore also adopt a power law for the clustering component in this work.

We model the total IR point source power spectrum from correlating maps at frequency $\nu_1$ and $\nu_2$ ($\nu_1=\nu_2$ for the auto-spectra) as
\begin{equation}
C_\ell(\nu_1,\nu_2)=A_{\textrm{c}}(\nu_1,\nu_2)\left(\frac{\ell}{\ell_0}\right)^{-n}+C_{\textrm{P}}(\nu_1,\nu_2),
\end{equation}
where $\ell$ is the multipole moment, $A_{\textrm{c}}$ and $n$ are the clustering amplitude and index, $C_{\textrm{P}}$ is the Poisson shot-noise and $\ell_0=3000$ is the pivot scale. This differs from the form adopted by \cite{amblard/etal:2011}  only in the choice of pivot $\ell_0$. Unlike P11 we model the clustering and shot-noise as separate, independent components rather than modeling their sum as one power law. Motivated by the apparent uniformity in angular scale dependence across the different spectra (see Figure 1) we fit for a single, frequency-independent, value of $n$. We fit for a separate shot-noise level in each of the 16 \emph{Planck}, BLAST and BLAST$\times$ACT spectra (see Section 3.2).

The SED of the CIB, over the range of frequencies considered, has been found to be well-described by a modified blackbody of the form \citep[e.g.,][]{fixsen/etal:1998,lagache/etal:1999,gispert/etal:2000}:
\begin{equation}
I_{\textrm{CIB}}(\nu)\propto\nu^{\beta}B(\nu,T_{\textrm{eff}}).
\end{equation}
In reality, the sources making up the CIB lie at a range of redshifts, with a variety of luminosities and dust temperatures \citep[e.g.,][]{haiman/knox:2000,knox/etal:2001,coppin/etal:2008,pascale/etal:2009,hwang/etal:2010}. Equation (2) is thus an approximation to the true CIB SED, which consists of the sum of many different (approximately) modified blackbody spectra, and the quantities $\beta$ and $T_{\textrm{eff}}$ are not to be interpreted as physical parameters.

P11 found that the SED of the CIB \emph{anisotropies} measured by \emph{Planck} is also well-described by the \cite{gispert/etal:2000} modified blackbody with emissivity spectral index $\beta=1.4\pm0.2$ and effective temperature $T_{\textrm{eff}}=13.6\pm1.5$~K. We assume the \emph{clustering power} SED can also be described in this form, and parameterise the frequency dependence of the auto-spectrum clustering power amplitude as $A_{\textrm{c}}=(I_{\textrm{c}})^2$ where
\begin{equation}
I_{\textrm{c}}(\nu)=I_{\textrm{0}}\left(\frac{\nu}{\nu_0}\right)^{\beta}B(\nu,T_{\textrm{eff}}),
\end{equation}
with a single emissivity index and effective temperature.

Different frequencies are sensitive to IR sources at different redshifts, with the importance of higher-redshift sources increasing at lower frequencies \citep[e.g.,][]{haiman/knox:2000,knox/etal:2001,chapin/etal:2009,marsden/etal:2009}. As a result we would not expect there to be 100\% correlation between the IR sources in different bands (particularly between the widely spaced ACT and BLAST bands) and we therefore include a clustering correlation factor $f_{\textrm{corr}}$ as a free parameter for each cross-spectrum. The cross-spectrum clustering amplitude is then $A_{\textrm{c}}(\nu_1,\nu_2)=f_{\textrm{corr}}(\nu_1,\nu_2)I_{\textrm{c}}(\nu_1)I_{\textrm{c}}(\nu_2)$.

\subsection{Shot-noise}

The auto-spectrum shot-noise is given by \citep[e.g.,][]{scott/etal:1999}
\begin{equation}
C_{\textrm{P}}=\int \textrm{d}z\int_0^{S_{\textrm{cut}}} \textrm{d}S \, S^2 \frac{\textrm{d}^2N}{\textrm{d}S\textrm{d}z}(S,z),
\end{equation}
where $S_{\textrm{cut}}$ is the flux cut applied to the map and $\textrm{d}^2N/\textrm{d}S\textrm{d}z$ are the differential source counts. 

In P11 the shot-noise levels were fixed using the IR galaxy evolution model of B\'ethermin et al. (2011 -- hereafter B11). This model parameterises the evolution of the galaxy luminosity function and is fit to number counts over a wide range of IR wavelengths. While the model broadly fits the available data, there are discrepancies. The model underpredicts by $\sim$40\% the shot-noise levels measured at 500~$\mu$m by BLAST \citep{viero/etal:2009} and 220~GHz by SPT \citep{hall/etal:2010}. We do not apply any priors on the shot-noise levels when constructing our template. We may, however, expect a considerable degeneracy between the shape of the clustering component and size of the shot-noise for the \emph{Planck} data, because \emph{Planck} is not able to probe the small scales where the shot-noise becomes dominant. As a result we consider the effect of adopting the B11 model predictions as priors on the \emph{Planck} shot-noise levels in Section 4.1.1.

In terms of the source flux and $S_{\textrm{cut}}$ we can write the clustering power as \citep[e.g.,][]{tegmark/etal:2002,viero/etal:2009}
\begin{equation}
C_{\ell}^{\textrm{clust}}=\int \textrm{d}z \left(\frac{\textrm{d}V}{\textrm{d}z}(z)\right)^{-1}\left(\frac{\textrm{d}S}{\textrm{d}z}(z)\right)^2P_{\textrm{gal}}(k,z)|_{k=\ell/\chi(z)},
\end{equation}
where $\chi$ is the comoving distance, $\textrm{d}V/\textrm{d}z=\chi^2\textrm{d}{\chi}/\textrm{d}z$ the comoving volume element, $P_{\textrm{gal}}$ the IR galaxy power spectrum, and the redshift-distribution of the flux, $\textrm{d}S/\textrm{d}z$, is given by
\begin{equation}
\frac{\textrm{d}S}{\textrm{d}z}(z)=\int_0^{S_{\textrm{cut}}} \textrm{d}S \, S\frac{\textrm{d}^2N}{\textrm{d}S\textrm{d}z}(S,z).
\end{equation}
The factor of $S^2$ in equation (4), compared to $(\textrm{d}S/\textrm{d}z)^2$ in equation (5), suggests that the removal of the highest-flux sources will have much less impact on the clustered power than on the shot-noise. The B11 model predicts that, for instance, applying a flux cut of 250~mJy (the cut applied to the BLAST 350~$\mu$m map in H12) to the \emph{Planck} 857~GHz map would result in a $\sim$10\% reduction in the shot-noise level compared to the \emph{Planck} flux cut of 710~ mJy, but that $(\textrm{d}S/\textrm{d}z)^2$ would be reduced by $<1$\% at $z\sim0.2$ and virtually unaffected for $z>1$. We allow for the dependence of the shot-noise levels on flux cut by fitting separate shot-noise levels for each spectrum, as described in the next section. We assume that the effect on the clustering power from applying different flux cuts is negligible for the data considered. Studies using higher-resolution \emph{Herschel} maps will be able to investigate the dependence of the clustering power on flux cut in more detail.

\subsection{MCMC fitting}

We perform a simultaneous fit to the seven auto-spectra (four \emph{Planck} and three BLAST) plus the nine cross-spectra (three BLAST $\times$ BLAST and six BLAST $\times$ ACT).  The model spectra are binned for comparison to each data spectrum, with likelihood
\be
-2\ln L(d|\theta) = \sum_{i=1}^{16} [{\bf C}_i^{\rm data}-{\bf C}^{\rm model}_i(\theta)]^2/{\bm \sigma}^2_i,
\ee
for model parameters $\theta$, data vector ${\bf C}_i^{\rm data}$ for the $i$th spectrum, and binned model spectra ${\bf C}^{\rm model}_i(\theta)$. Covariances between the sixteen spectra are neglected.

We find that the clustering SED parameters $I_{\rm 0}$, $\beta$ and $T_{\rm eff}$ (equation 3) are highly degenerate. Only two of these parameters are really independent and so we fix $T_{\rm eff}$ to the best-fit value, 9.7~K. We choose $\nu_0=530$~GHz to minimise the degeneracy between $I_{\rm 0}$ and $\beta$.

We therefore fit for 37 parameters: $n$, the clustering index, $I_0$, $\beta$, 16 shot-noise levels (one for each spectrum), nine cross-spectrum correlation factors (one for each cross-spectrum) and nine photometric calibration parameters (one for each \emph{Planck}, BLAST and ACT band). For each calibration factor we enforce a Gaussian prior centered at unity, with spread given by the nominal uncertainty listed in Section 2. The covariance between the BLAST calibration factors from \cite{truch/etal:2009} is included.  Apart from the calibration factors, all priors are uniform.

Given the high dimensionality, we estimate the posterior probability distribution using Markov Chain Monte Carlo (MCMC) analysis \citep{metropolis/etal:1953}, using the sampling methods and convergence test described in \cite{dunkley/etal:2005}. Chains used for analysis are about $10^6$ steps in length, and are used to calculate one-dimensional marginalized parameter values and errors.

To judge the goodness-of-fit of the model, we are also interested in the maximum likelihood. However, the peak of the likelihood distribution occupies only a small part of this high-dimensional parameter space, so the minimum $\chi^2$ sampled in a chain of about a million steps is significantly larger than the true value ($\Delta \chi^2 \sim 10$ for a simulated 37-dimensional Gaussian distribution). There are numerous statistical methods to find the true peak of a distribution. We adopt a simple modification to the Metropolis algorithm in which chains start from the best-fitting point and then make a step in parameter space only when the posterior is improved, using a reduced trial step size. We have tested this prescription on simulated Gaussian distributions, and the peak is found to within $\Delta \chi^2=0.1$ in $\sim 2\times10^4$ steps for 37 dimensions. We emphasize that this modified code is used only to assess the global goodness-of-fit, not to estimate the marginalized parameter distributions.

To estimate the model spectra at each frequency, it is not sufficient to use the nominal values of $\nu$ for the BLAST and ACT bands in equation (3), due to the finite width of each filter. The correct effective $\nu$ values depend on the SED of the emission mechanism. The clustering SED is estimated iteratively by repeating the MCMC fitting; each time the best-fit clustering SED is integrated through the BLAST and ACT filter profiles and an improved estimate of the effective $\nu$ values obtained. We found that after three iterations the SED converged, with effective frequencies of 1248, 829 and 607~GHz (effective wavelengths 240, 362 and 494~$\mu$m) for the nominal BLAST 250, 350 and 500~$\mu$m bands, and 220 and 150~GHz for the ACT 218 and 148~GHz bands. While the clustering SED is rising steeply at the ACT bands, the shift from nominal frequency is reduced by the narrowness of the ACT filters. For the \emph{Planck} data the nominal frequencies are used, because the temperature-to-flux conversion factors provided in P11 are given such that the flux is correct at the nominal frequencies.

\section{Results}

The model fits the data well. We find a best-fit $\chi^2$ of 132 for 122 degrees of freedom (150 data points minus 28 parameters; we have not counted the calibration parameters since they are strongly constrained by priors), giving reduced-$\chi^2=1.08$. The marginalized mean values of the clustered IR template parameters are given in Table 1. No errors are given for $T_{\rm eff}$, $\nu_0$ or $\ell_0$ since these parameters are fixed as described in Section 3.2. Figure 2 shows the best-fit clustered IR source power for each of the bands included in the fit. The best-fit shot-noise levels have been subtracted. The $\chi^2$ contribution from each individual spectrum is also shown; in addition to the total $\chi^2$ being good we find that there are no spectra which are not individually well-fit by the model. 

We plot the BLAST 350 and 500~$\mu$m spectra on the same axes as the \emph{Planck} 857 and 545~GHz spectra. The BLAST spectra are color-corrected to account for the difference in the bandpass filters by taking the ratio of our clustering power predictions for the relevant \emph{Planck} and BLAST band. We find these color correction factors to be 1.07 and 0.63 for the 350 and 500~$\mu$m BLAST spectra, respectively. P11 found values of 1.05 and 0.7 by integrating the SED of \cite{gispert/etal:2000} through the BLAST and \emph{Planck} filters. These values differ slightly because our best-fit clustering SED is slightly different -- see Figure 3 for a comparison.

\begin{figure*}[t]
	\centering
	\includegraphics{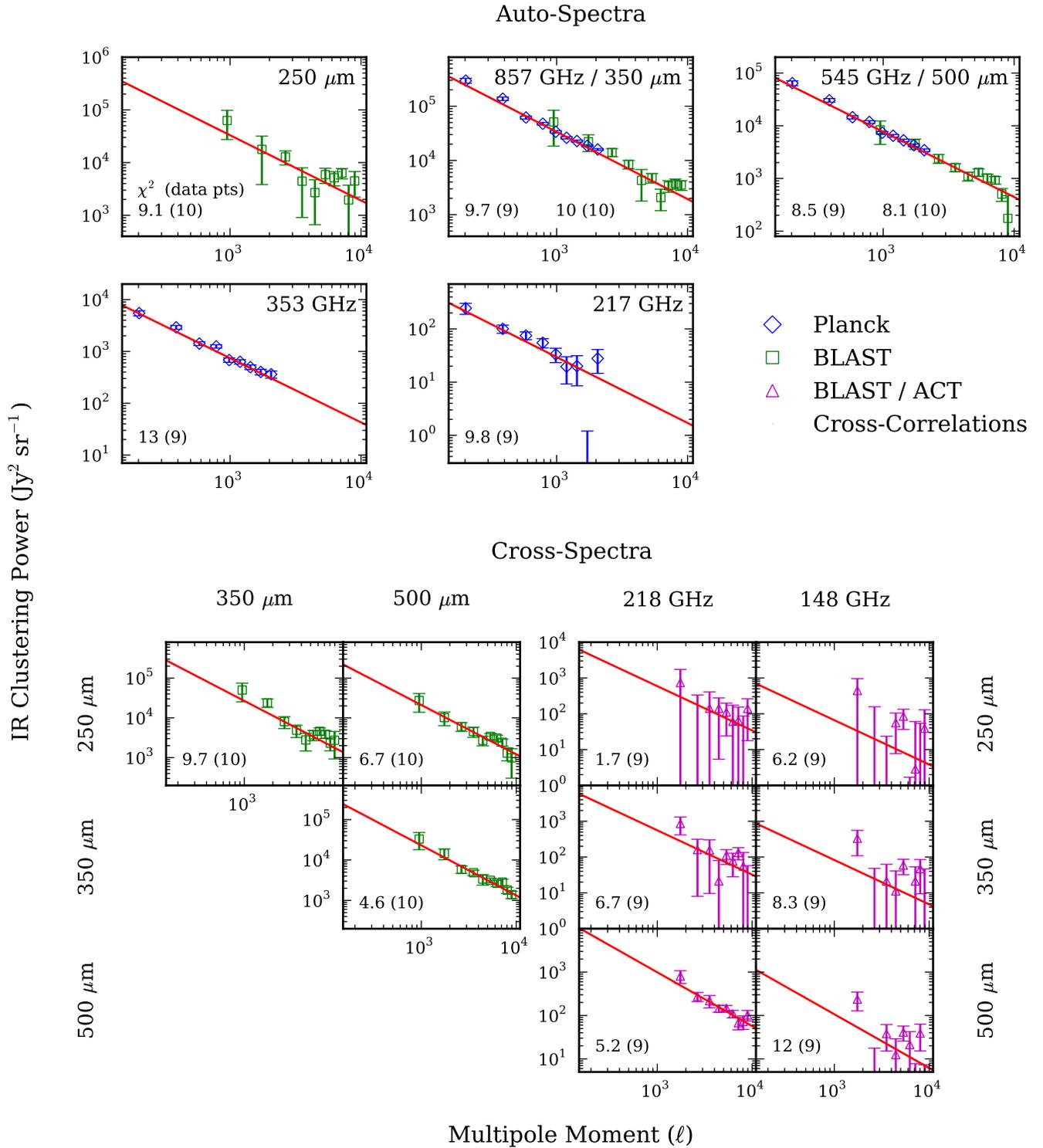}
	\caption{IR point source clustering power spectra from \emph{Planck} (diamonds), BLAST (squares) and BLAST / ACT cross-correlations (triangles). The solid lines are the best-fit power law with scale dependence $C_{\ell}^{\textrm{clust}}\propto\ell^{-n}$ -- we find $n=1.25\pm0.06$. The frequency dependence of the clustering SED is described by a modified blackbody (see equation 3 and Table 1). Poisson shot-noise has been subtracted for each panel (shot-noise levels from our fitting are given in Table 3). For the combined fit the $\chi^2/\textrm{degree of freedom}$ is $132/122$. The contribution to the total $\chi^2$ from the individual spectra is included in each panel. Color correction factors of 1.07 and 0.63 have been applied to the BLAST 350 and 500~$\mu$m spectra so they are directly comparable with the \emph{Planck} 857 and 545~GHz spectra, respectively, as described in the text.}
\end{figure*}

\begin{table}[h]
  \begin{center}
  \caption{Power-law clustering template}
  \begin{tabular}{lcc}
\hline
Parameter&Value\\
\hline
$n\tablenotemark{a}$
&$1.25\pm 0.06$\\
$I_{\textrm{0}}$ (Jy sr$^{-1}$)\tablenotemark{a}
&$(2.43\pm0.16)\times10^{-9}$\\
$\beta$
&$2.20\pm0.07$\\
$T_{\textrm{eff}}$ (K)
&$9.7$\\
$\nu_0$ (GHz)
&$530$\\
$\ell_0$
&3000\\
\hline
$\chi^2/\textrm{d.o.f.}$
&132/122\\
\hline
     \end{tabular}
     \tablenotetext{1}{A strong (89\%) anti-correlation exists between $n$ and $I_0$}
     \end{center}
\end{table}

The clustering power at $\ell=3000$ is shown as a function of frequency in Figure 3. Also shown is the scaled \cite{gispert/etal:2000} SED (dashed line), which was fit to FIRAS and DIRBE measurements of the CIB intensity spectrum. It falls off more slowly with decreasing frequency than our clustering SED. Some difference between the CIB mean and anisotropy SED may be expected, since the contribution of a source to the anisotropy SED depends on its clustering as well as spectral properties.

IR clustering power predictions at $\ell=2000$ and 3000, calculated from our template for several \emph{Planck}, ACT and SPT bands, are given in Table 2. The conversion from flux to temperature units was calculated in each case by integrating $I_{\textrm{c}}$ from equation (3) through the relevant bandpass filter. The effective frequencies (i.e., single frequency value that gives the same clustering amplitude as integrating through the filter) are also given. Since the IR clustering amplitude is rising strongly with frequency at the CMB bands (see Figure 3), the different filter profiles lead to significant differences in the clustering power even for bands with closely-spaced nominal frequencies. The amplitude of the clustered power may vary by a factor of 10 across a single filter, as shown in Figure 3.

\begin{table}[h]
  \centering
  \caption{IR clustering predictions from our template -- $\ell(\ell+1)C_{\ell}^{\textrm{clust}}/2\pi$ ($\mu$K$^2$)}
  \begin{tabular}{lcrclrcl}
\hline
Band&$\nu_{\textrm{eff}}$ (GHz)&\multicolumn{3}{c}{$\ell=2000$}&\multicolumn{3}{c}{$\ell=3000$}\\
\hline
\emph{Planck} 100~GHz
&104
&0.49\hspace*{-2mm}&$\pm$&\hspace*{-2mm}0.07
&0.67\hspace*{-2mm}&$\pm$&\hspace*{-2mm}0.09\\
\emph{Planck} 143~GHz
&146
&2.9\hspace*{-2mm}&$\pm$&\hspace*{-2mm}0.3
&3.9\hspace*{-2mm}&$\pm$&\hspace*{-2mm}0.4\\
\emph{Planck} 217~GHz
&226
&45\hspace*{-2mm}&$\pm$&\hspace*{-2mm}4
&61\hspace*{-2mm}&$\pm$&\hspace*{-2mm}6\\
\\
ACT 148~GHz
&150
&3.2\hspace*{-2mm}&$\pm$&\hspace*{-2mm}0.4
&4.4\hspace*{-2mm}&$\pm$&\hspace*{-2mm}0.5\\
ACT 218~GHz
&220
&36\hspace*{-2mm}&$\pm$&\hspace*{-2mm}3
&49\hspace*{-2mm}&$\pm$&\hspace*{-2mm}4\\
\\
SPT 95~GHz
&99
&0.38\hspace*{-2mm}&$\pm$&\hspace*{-2mm}0.05
&0.51\hspace*{-2mm}&$\pm$&\hspace*{-2mm}0.07\\
SPT 150~GHz
&156
&4.1\hspace*{-2mm}&$\pm$&\hspace*{-2mm}0.4
&5.5\hspace*{-2mm}&$\pm$&\hspace*{-2mm}0.6\\
SPT 220~GHz
&221
&37\hspace*{-2mm}&$\pm$&\hspace*{-2mm}3
&50\hspace*{-2mm}&$\pm$&\hspace*{-2mm}5\\
\hline
     \end{tabular}
\end{table}

The predictions at $\ell=3000$ can be compared to the ACT results from \cite{dunkley/etal:2011} and SPT results from \cite{shirokoff/etal:2011}. ACT find a best-fit clustering amplitude of $4.6\pm1.1$ and $54\pm13$~$\mu$K$^2$ at 148 and 218~GHz, respectively (we have calculated these uncertainties by adding their statistical and systematic error estimates in quadrature). SPT find $6.1\pm0.8$ and $57\pm8$~$\mu$K$^2$ at 150 and 220~GHz, respectively, for their base-line model, which includes a power-law IR clustering component with index $n=1.2$. Our predictions are in excellent agreement with these results. Figure 4 shows ACT and SPT data at 220~GHz with our clustering template predictions over-plotted, along with an IR shot-noise component and a $\Lambda$CDM CMB power spectrum.

We also make predictions for the spectral index $\alpha$ of the clustered component. At the CMB frequencies the clustering SED can be approximated as a power law, $I_{\textrm{c}}(\nu)\propto\nu^{\alpha}$, however, the CMB bands do not lie strictly in the Rayleigh-Jeans limit of the modified blackbody adopted in our model. Consequently the equivalent power-law slope at 150~GHz differs from that at 220~GHz: we find $\alpha_{150}=3.78\pm0.07$ and $\alpha_{220}=3.56\pm0.07$. We also calculate an effective spectral index between 150 and 220~GHz, $\alpha_{150-220}=3.68\pm0.07$. This result is consistent with ACT and SPT findings of $3.69\pm0.14$ and $3.58\pm0.08$, respectively. 

We have considered only the frequency dependence of the IR clustering component, not the shot-noise. Investigating shot-noise frequency dependence is more difficult, due to stronger flux cut dependence and the fact that the \emph{Planck} data are limited to large scales; future \emph{Herschel}, \emph{Planck}, ACT, SPT, and cross-correlation data will help overcome these issues.

\begin{figure*}[t]
	\hspace*{-6mm}
	\includegraphics{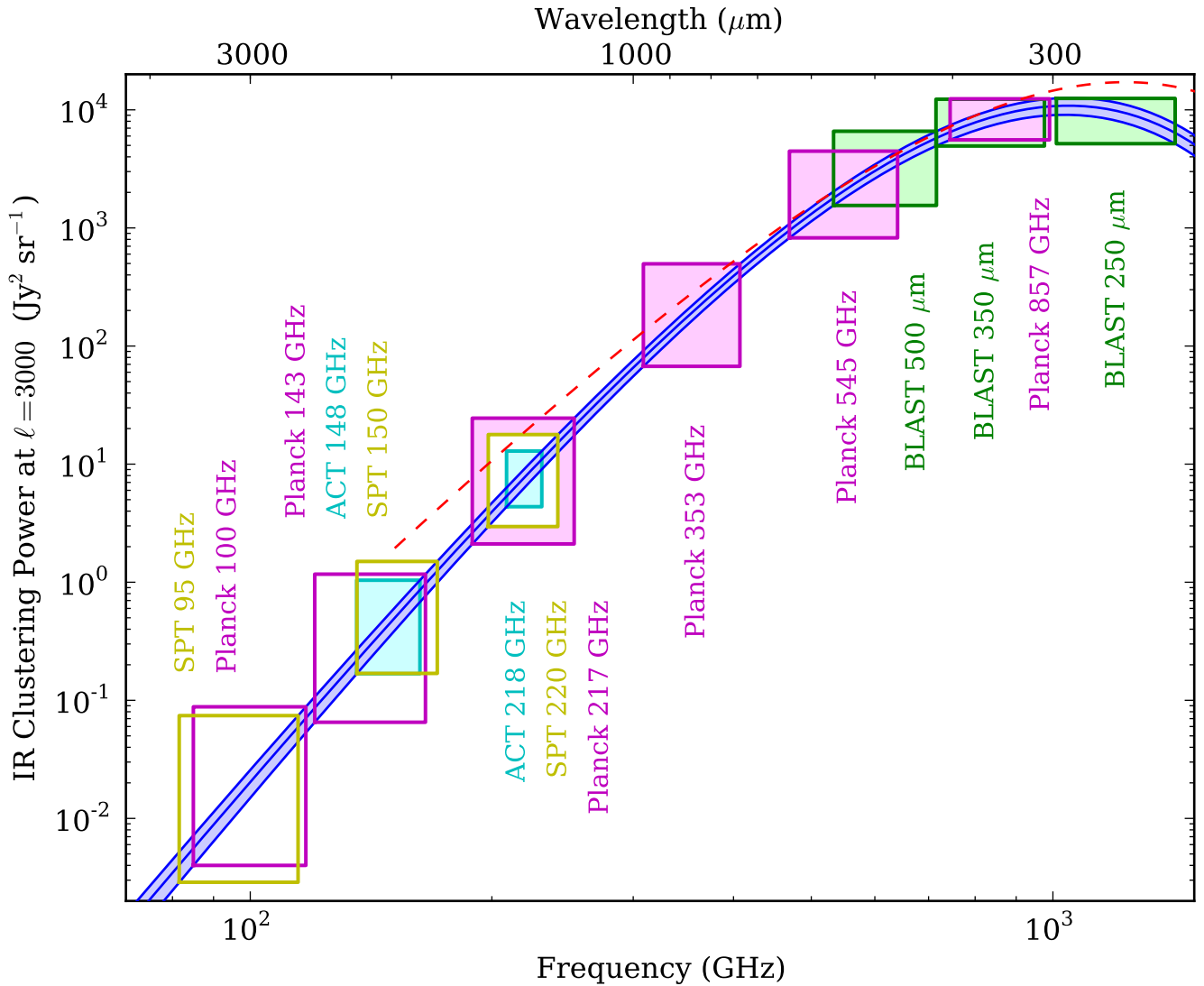}
	\caption{Frequency dependence of IR point source clustering power at $\ell=3000$. We assume this frequency dependence can be described by a modified blackbody: $C_{\ell=3000}^{\textrm{clust}}(\nu)\propto\left[\nu^{\beta}B(\nu,T_{\textrm{eff}})\right]^2$. The solid lines show the best-fit and 1-$\sigma$ uncertainties from our fitting; parameter values are given in Table 1. The dashed line is the SED of \cite{gispert/etal:2000}, which has been scaled for comparison. The rectangles show the \emph{Planck}, BLAST, ACT and SPT bandpass filters FWHM values in the horizontal direction. The vertical extent of the rectangles shows the variation of the clustering power across each filter; in some cases this variation is a factor of 10 or more due to how steeply the clustering SED rises with frequency. The shaded rectangles show the bands that were used for the fitting in this paper. Spectra from the other bands are either not yet available or were not used, due to the risk of biasing results by assumptions regarding the separation of the CMB and other components from the IR contribution.}
\end{figure*}

\subsection{Validating assumptions}

While our single-index power-law clustering model provides a good fit to the data in terms of $\chi^2$, in this section we attempt to further validate our assumptions regarding the scale and frequency dependence of the clustered power.

\subsubsection{Shot-noise}

Table 3 shows the 1-D marginalized values of the shot-noise levels from our MCMC chains ($C_\textrm{P}$ from equation 1). It should be noted that, as expected, the \emph{Planck} shot-noise levels and the clustering index $n$ are highly correlated. The shot-noise predictions for each auto-spectrum from the B11 model are given for comparison. Our values are consistent with the model within 1.5$\sigma$ in all cases. We repeated the MCMC analysis described in Section 3.2 using the B11 model predictions as Gaussian priors on the \emph{Planck} shot-noise levels and found a best-fit $\chi^2/\textrm{d.o.f.}=139/122$, an increase of $\Delta\chi^2=7$ compared to the case with uniform priors. There was minimal change in the clustering template parameters (e.g., $n=1.24\pm0.03$ with the shot-noise prior). This suggests that uncertainty in the \emph{Planck} shot-noise levels is not having a significant impact on our results.

\begin{table}[h]
  \centering
  \caption{1-D marginalized shot-noise levels}
  \begin{tabular}{lrclrcl}
\hline
Band&\multicolumn{3}{c}{Marginalized Value}&\multicolumn{3}{c}{B\'ethermin et al. (2011)}\\
&\multicolumn{3}{c}{$(\textrm{Jy}^2\textrm{ sr}^{-1})$}&\multicolumn{3}{c}{$(\textrm{Jy}^2\textrm{ sr}^{-1})$}\\
\hline
\emph{Planck} 857~GHz
&\hspace*{4mm}2200\hspace*{-2mm}&$\pm$&\hspace*{-2mm}2500
&\hspace*{7mm}5920\hspace*{-2mm}&$\pm$&\hspace*{-2mm}370\\
\emph{Planck} 545~GHz
&1700\hspace*{-2mm}&$\pm$&\hspace*{-2mm}700
&1150\hspace*{-2mm}&$\pm$&\hspace*{-2mm}90\\
\emph{Planck} 353~GHz
&210\hspace*{-2mm}&$\pm$&\hspace*{-2mm}60
&138\hspace*{-2mm}&$\pm$&\hspace*{-2mm}22\\
\emph{Planck} 217~GHz
&16\hspace*{-2mm}&$\pm$&\hspace*{-2mm}6
&12\hspace*{-2mm}&$\pm$&\hspace*{-2mm}3\\
BLAST 250~$\mu$m
&11600\hspace*{-2mm}&$\pm$&\hspace*{-2mm}2300
&11600\hspace*{-2mm}&$\pm$&\hspace*{-2mm}2100\\
BLAST 350~$\mu$m
&5200\hspace*{-2mm}&$\pm$&\hspace*{-2mm}1200
&5050\hspace*{-2mm}&$\pm$&\hspace*{-2mm}1080\\
BLAST 500~$\mu$m
&1300\hspace*{-2mm}&$\pm$&\hspace*{-2mm}390
&1680\hspace*{-2mm}&$\pm$&\hspace*{-2mm}480\\\\
250~$\mu$m~$\times$~350~$\mu$m
&5700\hspace*{-2mm}&$\pm$&\hspace*{-2mm}1400\\
250~$\mu$m~$\times$~500~$\mu$m
&1830\hspace*{-2mm}&$\pm$&\hspace*{-2mm}690\\
350~$\mu$m~$\times$~500~$\mu$m
&820\hspace*{-2mm}&$\pm$&\hspace*{-2mm}490\\
250~$\mu$m~$\times$~218~GHz
&240\hspace*{-2mm}&$\pm$&\hspace*{-2mm}130\\
250~$\mu$m~$\times$~148~GHz
&90\hspace*{-2mm}&$\pm$&\hspace*{-2mm}70\\
350~$\mu$m~$\times$~218~GHz
&240\hspace*{-2mm}&$\pm$&\hspace*{-2mm}70\\
350~$\mu$m~$\times$~148~GHz
&110\hspace*{-2mm}&$\pm$&\hspace*{-2mm}40\\
500~$\mu$m~$\times$~218~GHz
&66\hspace*{-2mm}&$\pm$&\hspace*{-2mm}30\\
500~$\mu$m~$\times$~148~GHz
&55\hspace*{-2mm}&$\pm$&\hspace*{-2mm}20\\
\hline
     \end{tabular}
\end{table}

\subsubsection{Photometric calibration}

The marginalized values for the photometric calibration parameters from our MCMC chains are shown in Table 4 along with the nominal uncertainties. We would expect the calibration parameters to be consistent with unity within the nominal uncertainty. If this were not the case it could indicate that the modified blackbody spectrum in equation (3) was not a suitable description for the frequency dependence of the clustering amplitude, however all are consistent with the nominal values.

\begin{table}[h]
  \centering
  \caption{1-D marginalized calibration parameters}
  \begin{tabular}{lcc}
\hline
Band&Marginalized Value&Nominal Uncertainty\\
\hline
\emph{Planck} 857~GHz
&1.05\hspace*{2mm}$\pm$\hspace*{2mm}0.06
&$7\%$\\
\emph{Planck} 545~GHz
&0.99\hspace*{2mm}$\pm$\hspace*{2mm}0.04
&$7\%$\\
\emph{Planck} 353~GHz
&1.01\hspace*{2mm}$\pm$\hspace*{2mm}0.02
&$2\%$\\
\emph{Planck} 217~GHz
&1.00\hspace*{2mm}$\pm$\hspace*{2mm}0.02
&$2\%$\\
BLAST 250~$\mu$m
&1.05\hspace*{2mm}$\pm$\hspace*{2mm}0.08
&$9.5\%$\\
BLAST 350~$\mu$m
&1.03\hspace*{2mm}$\pm$\hspace*{2mm}0.07
&$8.7\%$\\
BLAST 500~$\mu$m
&1.02\hspace*{2mm}$\pm$\hspace*{2mm}0.07
&$9.2\%$\\
ACT 218~GHz
&1.03\hspace*{2mm}$\pm$\hspace*{2mm}0.07
&$7\%$\\
ACT 148~GHz
&1.00\hspace*{2mm}$\pm$\hspace*{2mm}0.02
&$2\%$\\
\hline
     \end{tabular}
\end{table}

\subsubsection{\emph{Planck} beam uncertainty}

As stated in Section 2, we neglected any possible correlation in the \emph{Planck} beam uncertainty across different angular scales. We re-ran the MCMC chain enforcing a 100\% correlation in the beam uncertainty in different $\ell$-bins for each \emph{Planck} spectrum (introducing off-diagonal elements in the likelihood -- equation 7) and found minimal change in the marginalized parameter values ($<0.3\sigma$ in all cases). While the \emph{Planck} beam uncertainties at $\ell\sim2000$ are comparable to or larger than the statistical uncertainties, the template parameters are primarily constrained by the large-scale \emph{Planck} data (where the beam uncertainties are sub-dominant) and the BLAST data.

\subsubsection{Comparison with broken power law}

We assumed that the scale dependence of the clustering power can be described using a single index $n$. In this section we consider fitting the scale dependence with a broken power law of the form
\begin{equation}
C_{\ell}^{\textrm{clust}}\propto \left\{\begin{array}{rl}
&\ell^{-n_1} \mbox{  if $\ell\le\ell_{\rm b}$} \\
&\ell^{-n_2} \mbox{  if $\ell>\ell_{\rm b}$,}
  	\end{array} \right.
\end{equation}
where $n_1$, $n_2$ and $\ell_b$ are free parameters. When we repeat the fit to the \emph{Planck}, BLAST and BLAST $\times$ ACT spectra with this new scale dependence we find $n_1=1.24\pm0.06$, $n_2=1.2\pm0.2$ and $\ell_b=1100\pm300$ with a best $\chi^2/\textrm{d.o.f.}=131/120$. This is an improvement of only $\Delta\chi^2=0.9$ with two additional fitted parameters. For the fit with priors on the \emph{Planck} shot-noise levels from the B11 model the improvement is only $\Delta\chi^2=0.4$. We conclude that the data do not show a significant preference for a broken power law.

\cite{keisler/etal:2011} adopted a template of the form given in equation (8) with $n_1=2.0$, $n_2=1.2$ and $\ell_{\textrm b}=1500$ to model the IR clustered power when extracting cosmological information from the small-scale SPT 150 GHz CMB power spectrum. We find that this form is not a good fit to the \emph{Planck} data, with $\chi^2/\textrm{d.o.f.}=22/7$ when we fit to the \emph{Planck} 217~GHz spectrum with $n_1$, $n_2$ and $\ell_{\textrm b}$ fixed to the above values and using uniform priors on the clustering amplitude and shot-noise.

\subsubsection{Comparison with linear bias model}

H12 found that the BLAST and BLAST $\times$ ACT data are well-fit by assuming the IR galaxy power spectrum in equation (5) is given by
\begin{equation}
P_{\textrm{gal}}(k,z)=b^2P_{\textrm{DM}}(k,z),
\end{equation}
where $b$ is the linear bias factor and $P_{\textrm{DM}}$ is the linear dark matter power spectrum, and using the B11 predictions for the redshift-distribution of the flux, $\textrm{d}S/\textrm{d}z$. P11 also found that the \emph{Planck} data from each band are well-fit by this model (also using the B11 $\textrm{d}S/\textrm{d}z$) if no prior is enforced on the shot-noise levels. The bias levels are not consistent; when fitting for a single, redshift-independent value of $b$, H12 found $b=5.0\pm0.4$ whereas P11 found $b=2.18\pm0.11$ for the \emph{Planck} 545~GHz spectrum.

The linear bias model is strongly rejected by the combined \emph{Planck} and BLAST data even without any shot-noise prior. Fitting a linear bias model with single-value bias to the \emph{Planck} 857~GHz and BLAST 350~$\mu$m spectra together, using the B11 $\textrm{d}S/\textrm{d}z$ and multiplying the BLAST data points by a color correction factor of 1.07 (see Section 4), yields $\chi^2/\textrm{d.o.f.}=39/16$. For a joint fit to the \emph{Planck} 545~GHz and BLAST 500~$\mu$m spectra we find $\chi^2/\textrm{d.o.f.}=43/16$. This result is driven by the shape of the linear matter power spectrum rather than the choice of $\textrm{d}S/\textrm{d}z$: we repeated the fitting using the predictions of \cite{marsden/etal:2011} rather than B11 and found $\chi^2/\textrm{d.o.f.}$ of 36/16 and 42/16 for the 857~GHz / 350~$\mu$m and 545~GHz / 500~$\mu$m spectra respectively. Neither the \emph{Planck} nor BLAST data alone were able to rule out the linear bias model without a shot-noise prior because of the limited angular scales probed. \cite{shirokoff/etal:2011} similarly found that the linear bias model could not be ruled out using only SPT data from $\ell\gtrsim2000$.

\begin{figure*}[t]
	\includegraphics{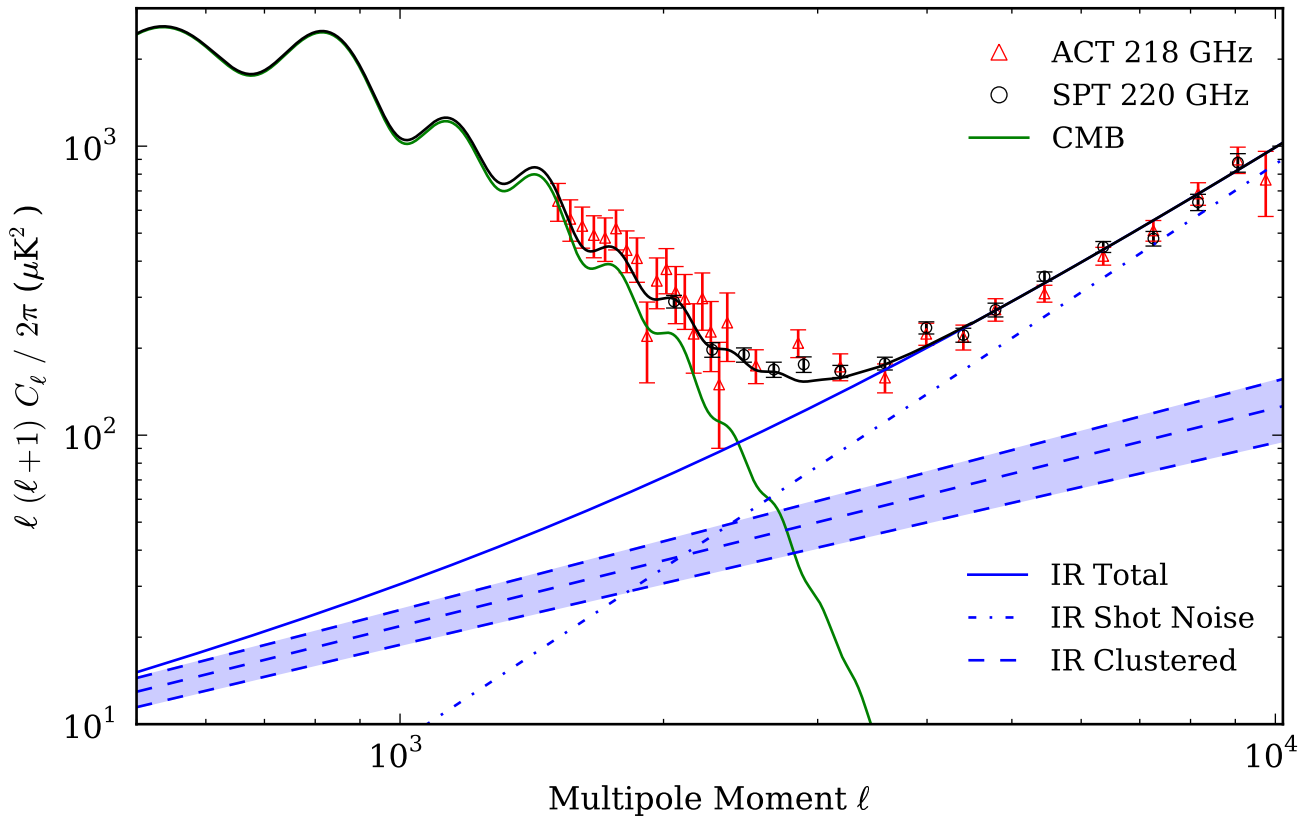}
	\caption{The angular power spectra at 220~GHz measured by ACT \citep{das/etal:2011} and SPT \citep{shirokoff/etal:2011}, with a theoretical model for CMB and infrared point sources over-plotted. The clustered IR component is given by the power-law model fit to \emph{Planck}, BLAST and BLAST $\times$ ACT data in this analysis (see equations 1 and 3 and Table 1 for parameter values). The upper and lower dashed lines correspond to 1$\sigma$ error bounds. The lensed CMB spectrum is that of the $\Lambda$CDM model with parameters derived from \WMAP\ \citep{komatsu/etal:2011}. An IR shot-noise component of size $\ell(\ell+1)C_{\textrm{P}}/2\pi |_{\ell=3000}=78$~$\mu$K$^2$ (consistent with ACT and SPT measurements) has also been plotted. Additional radio source and kinetic SZ power are subdominant at this frequency and are not included.}
\end{figure*}

\subsection{Cross-correlations}

Table 5 shows the marginalized degree of clustering correlation, $f_{\textrm{corr}}$, for each cross-spectrum included in our fitting, along with the separation $\Delta\nu$ between the two bands. The BLAST $\times$ BLAST cross-spectra are consistent with 100\% correlation. Our conclusions regarding the degrees of correlation between the ACT and BLAST bands are limited by the data quality. A decrease in correlation with increasing band separation would be consistent with the sources lying at a range of redshifts, with the higher-redshift sources being of greater relative importance at the longer wavelengths \citep[e.g.,][]{haiman/knox:2000}, however the current data are not of sufficient quality to confirm this.

Two of the marginalized mean $f_{\rm corr}$ values lie more than 1$\sigma$ above unity, while none lie more than 1$\sigma$ below. Measuring $f_{\rm corr}>1$ may indicate that the angular scale dependence of the cross-spectra clustering power is not described by the same single-index power law as the auto-spectra clustering, since there is no physical explanation for a correlation in excess of 100\%. This could be the case even with no worsening of the $\chi^2$, due to the limited angular scales probed by the BLAST $\times$ ACT data. To test that this is not having a significant effect on our clustering template, we repeated the MCMC fitting described in Section 3.2 using only the auto-spectrum data. We found that the values of $n$, $\beta$ and $I_0$ change by $<0.5\sigma$ compared to the fit with the cross-spectra included. We conclude that, while the data may be hinting that the cross-spectrum clustering power has a different shape to the auto-spectrum clustering, this is not significantly biasing our template.

Further submm-mm cross-correlation studies (e.g., \emph{Herschel} $\times$ ACT / SPT) are clearly required to provide more insight into the distribution of the sources with redshift, to constrain the angular scale dependence of the cross-spectrum clustering, and to investigate how the clustering shape changes with increasing band separation.

\begin{table}[h]
  \centering
  \caption{Degree of clustering cross-correlation, $f_{\textrm{corr}}$}
  \begin{tabular}{cccc}
\hline
Band 1&Band 2&$\Delta\nu$ (GHz)&Marginalized Value\\
\hline
250~$\mu$m
&350~$\mu$m
&340
&0.98\hspace*{2mm}$\pm$\hspace*{2mm}0.30\\
250~$\mu$m
&500~$\mu$m
&600
&1.23\hspace*{2mm}$\pm$\hspace*{2mm}0.31\\
350~$\mu$m
&500~$\mu$m
&260
&1.41\hspace*{2mm}$\pm$\hspace*{2mm}0.28\\
250~$\mu$m
&218~GHz
&982
&0.66\hspace*{2mm}$\pm$\hspace*{2mm}1.09\\
250~$\mu$m
&148~GHz
&1052
&0.30\hspace*{2mm}$\pm$\hspace*{2mm}1.83\\
350~$\mu$m
&218~GHz
&642
&0.64\hspace*{2mm}$\pm$\hspace*{2mm}0.56\\
350~$\mu$m
&148~GHz
&712
&0.39\hspace*{2mm}$\pm$\hspace*{2mm}1.05\\
500~$\mu$m
&218~GHz
&382
&1.82\hspace*{2mm}$\pm$\hspace*{2mm}0.47\\
500~$\mu$m
&148~GHz
&452
&0.79\hspace*{2mm}$\pm$\hspace*{2mm}0.87\\
\hline
\end{tabular}
\end{table}

\section{Conclusions}

We have found that a power-law model for the IR point source clustering is adequate to simultaneously fit \emph{Planck}, BLAST and cross-correlated BLAST / ACT power spectrum data over a broad range of frequency ($150<\nu<1200$~GHz) and angular scale (multipole moment $100<\ell<9000$). We find the clustering power varies with angular scale as $\ell^{-n}$ with $n=1.25\pm0.06$ and that the SED of the clustering can be described as a modified blackbody with emissivity index $\beta=2.20\pm0.07$ and effective temperature $T_{\rm eff}=9.7$~K. Our work does not rely on any assumptions regarding the physical properties of the IR sources (host halos, redshift distribution, etc.).

As well as providing a simple template for use in CMB foreground subtraction, we have established that the \emph{Planck} and BLAST / BLAST $\times$ ACT data sets appear compatible when bandpass filters, flux cut and calibration are accounted for, as described in Sections 2 and 3. We make predictions for the IR clustering power for the \emph{Planck}, ACT and SPT CMB bands; our predictions for the ACT and SPT bands at around 150 and 220~GHz are fully consistent with existing measurements \citep{dunkley/etal:2011,shirokoff/etal:2011}.

There is uncertainty in the \emph{Planck} shot-noise levels because \emph{Planck} does not probe small enough scales for the shot-noise to become dominant. We find reasonable consistency between the shot-noise levels from our fitting and the predictions of the parametric IR galaxy evolution model of \cite{bethermin/etal:2011}. We repeated our fitting using this model's predictions as priors on the \emph{Planck} shot-noise levels and found there was minimal effect on the scale or frequency dependence of the clustering power. Number counts obtained via $P(D)$ analysis (probability of deflection -- modeling the probability distribution function of observed flux in each map pixel -- e.g., \citealt{patanchon/etal:2009}; \citealt{glenn/etal:2010}) may be useful for constraining shot-noise levels in future work.

Upcoming data from \emph{Herschel}, \emph{Planck}, ACT, SPT, and cross-correlations will allow us to look for any variations in the clustering angular scale dependence with frequency. Understanding this scale dependence in the mm CMB bands is important not only for extracting unbiased estimates of cosmological parameters, but also for constraining other components of the measured spectrum, such as the thermal and kinetic SZ effect, and any cross-component correlations, for instance SZ-IR.

Over the range of angular scales considered we would expect to be probing both the linear and non-linear clustering regimes. This is supported by the fact that the combined \emph{Planck} and BLAST data strongly reject a linear bias clustering model (Section 4.1.5). The linear and non-linear components have different scale dependence, with the linear component dominating on large scales and the non-linear on small scales. The fact that our study has found that a power-law scale dependence provides a good fit to the \emph{Planck} and BLAST IR point source clustering power is not inconsistent with this picture; instead it suggests that the sum of the linear and non-linear components is sufficiently close to a power law that the current data are unable to reveal any deviations. We can draw a comparison to measurements of the angular correlation function, $w(\theta)$, of luminous red galaxies and galaxies in the main SDSS sample, which have revealed deviations from power law behavior \citep[e.g.,][]{zehavi/etal:2004,zehavi/etal:2005,blake/etal:2008} whereas earlier studies of galaxy clustering were unable to do so \citep[e.g.,][]{zehavi/etal:2002}. We may expect similar deviations to be detected in future IR source power spectra, in particular with \emph{Herschel}/SPIRE data, which will yield higher quality data than BLAST out to smaller scales with its improved angular resolution.

The correlation function of resolved sub-mm sources \citep[e.g.,][]{cooray/etal:2010,maddox/etal:2010} provides complementary information to the power spectra of unresolved sources considered in this work. Current data is limited; \cite{guo/etal:2011} measure $\xi(r)\propto r^{-\gamma}$ with $\gamma\sim2$, corresponding to $C_{\ell}^{\rm clust}\propto\ell^{-1}$, for low-redshift ($z<0.5$) \emph{Herschel}-ATLAS galaxies, although their result is consistent with ours within errors. The full \emph{Herschel}-ATLAS survey \citep{eales/etal:2010} will cover 30 times more sky and provide far tighter constraints. Discrepancies between the $C_{\ell}^{\rm clust}$ and $\xi(r)$ or $w(\theta)$ measurements would indicate differences in clustering properties of the bright sub-mm galaxies compared to the fainter population, and thus both statistics are important for constraining models of galaxy clustering and evolution.

\setlength{\parskip}{2ex}

This work was supported by the U.S. National Science Foundation through awards AST-0408698 for the ACT project, and PHY-0355328, AST-0707731 and PIRE-0507768. GA is supported by an STFC studentship and funding was also provided by Princeton University and the University of Pennsylvania, RCUK Fellowship (JD), ERC grant 259505 (JD) and NASA grant NNX08AH30G (AH). ACT operates in the Chajnantor Science Preserve in northern Chile under the auspices of the Comisi\'on Nacional de Investigaci\'on Cient\'ifica y Tecnol\'ogica (CONICYT). Data acquisition electronics were developed with assistance from the Canada Foundation for Innovation. We thank Matthieu B\'ethermin for discussions about the B11 IR model, Olivier Dor\'e, Guilaine Lagache and Bill Jones for discussions about the \emph{Planck} data, Joaquin Vieira for information about SPT filters, and Gaelen Marsden, Bruce Partridge and George Efstathiou for helpful suggestions.

\end{document}